\documentclass[]{spie} 
\renewcommand{\baselinestretch}{1.0} 

\usepackage[pdftex]{graphicx}  
\usepackage{float}

\usepackage{amsmath,amsfonts,amssymb}

\usepackage[colorlinks=true, allcolors=blue]{hyperref}

\usepackage{multirow}

\usepackage{xcolor}
\usepackage{verbatim}  

\DeclareUnicodeCharacter{FF0C}{,}      
\DeclareUnicodeCharacter{3002}{.}      
\DeclareUnicodeCharacter{FF1A}{:}      
\DeclareUnicodeCharacter{FF1B}{;}      
\DeclareUnicodeCharacter{FF01}{!}      
\DeclareUnicodeCharacter{FF1F}{?}      
\DeclareUnicodeCharacter{FF08}{(}      
\DeclareUnicodeCharacter{FF09}{)}      
\DeclareUnicodeCharacter{201C}{``}     
\DeclareUnicodeCharacter{201D}{''}     
\DeclareUnicodeCharacter{2018}{`}      
\DeclareUnicodeCharacter{2019}{'}      
\DeclareUnicodeCharacter{3001}{,}      
\DeclareUnicodeCharacter{2026}{\ldots} 
\DeclareUnicodeCharacter{2014}{---}    
\DeclareUnicodeCharacter{00A0}{~}      

\begin{document}

\pagestyle{plain}
\setcounter{page}{1} 

\title{ADNP-15: An Open-Source Histopathological Dataset for Neuritic Plaque Segmentation in Human Brain Whole Slide Images with Frequency Domain Image Enhancement for Stain Normalization}
\author[a]{Chenxi Zhao}
\author[a]{Jianqiang Li}
\author[a]{Qing Zhao}
\author[a]{Jing Bai}
\author[d]{Susana Boluda}
\author[d]{Benoit Delatour}
\author[d]{Lev Stimmer}
\author[b]{Daniel Racoceanu}
\author[c]{Gabriel Jimenez\textsuperscript{*}}
\author[a]{Guanghui Fu\textsuperscript{*}}

\affil[a]{School of Software Engineering, Beijing University of Technology, Beijing, China}
\affil[b]{Sorbonne Université, Institut du Cerveau - Paris Brain Institute - ICM, CNRS, Inria, Inserm, AP-HP, Hôpital de la Pitié Salpêtrière, Paris, France}
\affil[c]{Sorbonne Université, Institut du Cerveau - Paris Brain Institute
- ICM, Inserm, CNRS, APHP, Hôpital de la Pitié Salpêtrière, Paris, France}
\affil[d]{Sorbonne Université, Institut du Cerveau - Paris Brain Institute
- ICM, Inserm, CNRS, APHP, Hôpital de la Pitié Salpêtrière, DMU Neuroscience, Paris, France}

\authorinfo{Further author information: (Send correspondence to Gabriel Jimenez and Guanghui Fu, first.last@icm-institute.org)}

\maketitle

\begin{abstract}
Alzheimer’s Disease (AD) is a neurodegenerative disorder characterized by amyloid-$\beta$ plaques and tau neurofibrillary tangles, which serve as key histopathological fiatures. The identification and segmentation of these lesions are crucial for understanding AD progression but remain challenging due to the lack of large-scale annotated datasets and the impact of staining variations on automated image analysis. Deep learning has emerged as a powerful tool for pathology image segmentation; however, model performance is significantly influenced by variations in staining characteristics, necessitating effective stain normalization and enhancement techniques.
In this study, we address these challenges by introducing an open-source dataset (ADNP-15) of neuritic plaques (i.e., amyloid deposits combined with a crown of dystrophic tau-positive neurites) in human brain whole slide images. We establish a comprehensive benchmark by evaluating five widely adopted deep learning models across four stain normalization techniques, providing deeper insights into their influence on neuritic plaque segmentation. Additionally, we propose a novel image enhancement method that improves segmentation accuracy, particularly in complex tissue structures, by enhancing structural details and mitigating staining inconsistencies. Our experimental results demonstrate that this enhancement strategy significantly boosts model generalization and segmentation accuracy.
All datasets and code are open-source, ensuring transparency and reproducibility while enabling further advancements in the field. 

\end{abstract}


\section{Introduction} \label{sec:intro}
Alzheimer's Disease (AD) is a complex and prevalent neurodegenerative disorder characterized by distinct brain lesions, including amyloid-$\beta$ plaques and tau neurofibrillary tangles, which are crucial biomarkers in its diagnosis and progression\cite{montine2012national}
These lesions vary significantly in their morphological and topographical presentation, contributing to the clinical and pathological heterogeneity of the disease. Current diagnostic frameworks, such as the ABC scoring system, offer standardized methods for assessing AD pathology but are limited in capturing these lesions' nuanced variations and fine-grained heterogeneity\cite{shanthi2013image}.
Therefore, advanced methodologies combining digital pathology, artificial intelligence, and spatial morphological analysis have been proposed to refine our understanding of AD and its diverse clinical manifestations~\cite{Border2022}.
A significant challenge in these methodologies is the identification and quantification of amyloid-$\beta$ plaques and tau tangles, typically achieved through manual methods or semi-automated proprietary software\cite{benveniste1999detection}
This process can be labor-intensive and prone to variability due to human observation and expertise. Additionally, the accuracy of these analyses can be affected by variations in histological slide staining. To address these issues, implementing color normalization techniques is essential for standardizing digital pathology workflows and ensuring reliable automated image analysis by leveraging these innovations. Researchers aim to enhance AD pathology's diagnostic precision and understanding, ultimately contributing to improved patient outcomes.

Deep learning methods have made significant progress in this field, providing powerful tools for automatically detecting and segmenting pathological features in AD~\cite{salvi2021impact}. These methods offer the potential to overcome the limitations of traditional manual and semi-automated methods by achieving more accurate and reproducible analysis. For instance, convolutional neural networks (CNNs) such as UNet~\cite{ronneberger2015u} and nnUNet~\cite{isensee2021nnu} have demonstrated strong performance in medical image segmentation tasks. However, deep learning in AD pathological image segmentation faces two major challenges.
The first challenge is the lack of large-scale annotated datasets, which limits the generalization ability of models. Many existing studies rely on small datasets; for example, Wurts et al.\cite{wurts2020segmentation} used data from a single patient, and Maňoušková et al.\cite{mavnouvskova2022tau} used six subjects. In contrast, our study includes data from 15 subjects, providing a more comprehensive dataset for model training and evaluation.
The second challenge is the influence of staining variations, which significantly impact deep learning model performance~\cite{hoque2024stain}. Stain heterogeneity~\cite{ke2021style} introduces inconsistencies in model inputs, making segmentation less robust. The underlying mechanisms behind these effects require further investigation, and stain normalization has been recognized as an effective approach to mitigate these variations. To address this, researchers have explored image preprocessing techniques such as color normalization and image enhancement to improve model robustness and accuracy~\cite{tellez2019quantifying, salvi2024generative, tam2016method}.

In this study, we build upon our previous work presented at MICCAI 2022~\cite{jimenez2022visual}, where we explored the impact of color normalization on deep learning models for neuritic plaque segmentation in whole slide images of human brain tissue. A key advancement in this work is the release of an open-source dataset named as ADNP-15, including both the raw data and images processed with four different stain normalization methods, as well as augmented versions. This significantly extends the MICCAI version, which only provided images processed with two stain normalization methods. Additionally, we establish a comprehensive benchmark by evaluating five widely adopted deep learning models across the four stain normalization techniques, offering deeper insights into their impact on segmentation performance. To further improve segmentation accuracy, we propose a novel image enhancement method that enhances structural details and mitigates staining inconsistencies. Experimental results demonstrate that this enhancement method consistently improves segmentation accuracy, particularly in handling complex tissue structures. All datasets, methods, and code are fully open-source, ensuring transparency and reproducibility while facilitating further advancements in this field.


\section{Related Work}

In recent years, deep learning techniques have demonstrated significant potential in the pathological image analysis of Alzheimer's disease, particularly in detecting and quantifying tau protein, a key histopathological marker strongly associated with clinical symptoms. However, the specific mechanism of action of this marker in AD is still not fully understood.

Maňoušková et al.~\cite{mavnouvskova2022tau} proposed a deep learning-based histopathology pipeline for detecting and segmenting tau protein aggregates in Alzheimer's disease patients. Applied to six ALZ50-immunostained annotated brain tissue slices, their method used nonlinear color normalization, and a CNN-U-Net-based classification/segmentation model. They reported detection F1-scores of 75.8\% for neurofibrillary tangles and 81.3\% for neuritic plaques, with segmentation F1-scores of 91.1\% and 78.2\%, respectively. This study enables automated tau protein segmentation in Alzheimer's disease, aiding in clinical pathology analysis and subtype classification. However, they used the F1-score as the evaluation metric for both detection and segmentation tasks, which may not be entirely appropriate. In segmentation tasks, the Dice score is likely a more informative metric, as the F1-score relies on a threshold for overlap, potentially introducing bias. This concern is particularly relevant because their approach uses the CNN classification output to identify potential object regions before segmentation, yet they neither tested nor reported the performance of the complete pipeline, limiting the validity of their evaluation.
Wurts et al.~\cite{wurts2020segmentation} evaluated FCN, SegNet, and U-Net for tau pathology segmentation in Alzheimer's brain tissue using noisy labels. Trained on FFPE (Formalin fixed, paraffin embedded) tissue sections from a single patient with hematoxylin and DAB staining, the models achieved Dice scores of 0.234 (SegNet), 0.297 (FCN), and 0.272 (U-Net), demonstrating their ability to segment tau pathology despite noisy labeling.
Ingrassia et al.~\cite{ingrassia:hal-04793579} developed a deep learning workflow for annotating and segmenting neuritic plaques (NPs) and neurofibrillary tangles (NFTs) in Alzheimer's brain sections. Applied to 15 AT8-immunostained prefrontal cortex sections from four different biobanks, their AI-driven iterative approach enhanced annotation quality. Two U-Net models achieved Dice scores of 0.77 (NPs) and 0.81 (NFTs), demonstrating high accuracy and consistency. However, despite these promising results, the reproducibility of their approach is limited, as the methods were implemented using the proprietary software Visiopharm, making it difficult for independent validation and replication.
Ghandian et al.~\cite{ghandian2024learning} proposed a deep learning method for segmenting neurofibrillary tangles (NFTs) in Alzheimer's brain tissue from single-point annotations. Applied to 15 full-length slices from three research centers, their approach generated pixel-level segmentation masks from point annotations and trained a U-Net model, achieving a precision of 0.53, recall of 0.60, and F1-score of 0.53 on a test set.
Signaevsky et al.~\cite{signaevsky2019artificial} developed a deep learning method for assessing neurofibrillary tangles (NFTs) in Alzheimer's and other tau pathologies. Using WSI from 22 autopsy brains, they trained an FCN to quantify tau burden with high accuracy. The model achieved a recall of 0.92, precision of 0.72, and F1-score of 0.81, demonstrating its effectiveness for large-scale NFT detection.

From a data perspective, our dataset is larger than most related studies, allowing for a more comprehensive assessment of variability in the staining procedure and its impact on deep learning approaches. For instance, Wurts et al.\cite{wurts2020segmentation} used data from a single patient, and Maňoušková et al.\cite{mavnouvskova2022tau} used six patients, whereas our study includes data from 15 subjects. Additionally, we openly publish our dataset to support further research in this field.
In terms of normalization methods, our study evaluated a broader range of techniques. While Maňoušková et al.\cite{mavnouvskova2022tau} and Jiménez et al.\cite{jimenez2023meta} applied various color normalization methods, including Macenko, Reinhard, and Vahadane, other studies, such as Signaevsky et al.\cite{signaevsky2019artificial}, did not incorporate color normalization in their pre-processing steps. Our study tested four normalization methods—Reinhard, Vahadane, Macenko, and HistomicsTK—offering a more comprehensive evaluation.
Regarding model validation, we tested five deep learning models, whereas Maňoušková et al.\cite{mavnouvskova2022tau} used only the U-Net model, and Wurts et al.\cite{wurts2020segmentation} tested three models. This broader evaluation enhances the robustness of our findings.
For performance evaluation, most studies rely primarily on the Dice score. In contrast, we assess segmentation accuracy from multiple perspectives, incorporating Dice score, Mean Absolute Surface Distance (MASD), and instance-level F1-score. Our findings also highlight the challenges of this task, as some methods struggle to achieve high performance; for example, Wurts et al.~\cite{wurts2020segmentation} reported a Dice score of approximately 0.2. Our data augmentation approach significantly improves model performance, demonstrating its effectiveness.
Overall, our contributions enhance both the practicality and accuracy of segmentation models while providing a reliable foundation and valuable resources for future research.

\section{Methods}

\subsection{Frequency domain image enhancement}
\begin{figure}[!hbtp]
\centering\includegraphics[width=1.0\textwidth]{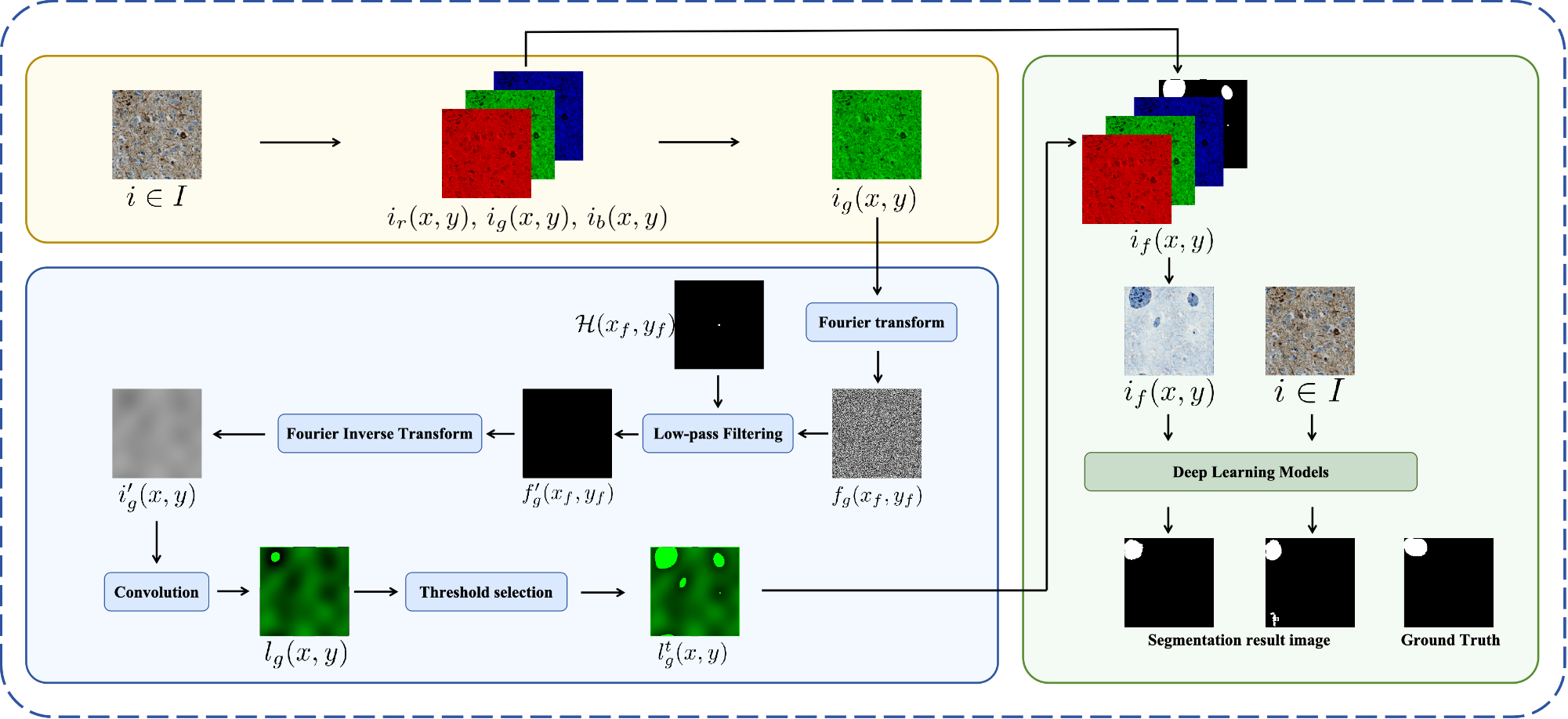}
\caption{The architecture of the proposed method.}
\label{fig:method}
\end{figure}

We proposed a simple image enhancement method based on the Fourier transform, as illustrated in Figure~\ref{fig:method}. This operation generates a novel representation that emphasizes the target region, potentially enhancing model performance.

Given sample $i \in I$, where $I \subset \mathbb{R}^{N_x \times N_y \times 3}$ represents a set of images, we decompose each image $i$ into three color channels: red, green, and blue (RGB), denoted as $i_r(x, y)$, $i_g(x, y)$, and $i_b(x, y)$, respectively.
We specifically choose the green channel $i_g$, which typically provides great detail and texture information from the image, to perform the following operation. First, we apply the Fourier transform $\mathcal{FT}$ to transfer $i_g$ into the frequency domain, obtaining the frequency representation $f_g(x_f, y_f)$ shown in Equation~\ref{eq:fourier}. 
\begin{equation}\label{eq:fourier}
\begin{aligned}
f_g(x_f, y_f)=\mathcal{FT}({i_g}(x,y)) \\
\end{aligned}
\end{equation}
Next, we apply a low-pass filter $\mathcal{H}$, where $c_f$ denotes the cutoff frequency shown in Equation~\ref{eq:low_pass}. 
The filtered frequency domain representation is $f_g'(x_f, y_f)$, and $\mathcal{H}(x_f, y_f)$ is the low-pass filter applied in the frequency domain.  
\begin{equation}\label{eq:low_pass}
\begin{aligned}
\mathcal{H}(x_f, y_f) =
\begin{cases} 
1, & \sqrt{x_f^2 + y_f^2} \leq c_f, \\
0, & \sqrt{x_f^2 + y_f^2} > c_f. \\
\end{cases} \\
f_g'(x_f, y_f) = f_g(x_f, y_f) \times \mathcal{H}(x_f, y_f) \\
\end{aligned}
\end{equation}

We apply the inverse Fourier transform, $\mathcal{FT}^{-1}$, to convert the filtered frequency representation back into the image space, as shown in Equation~\ref{eq:inverse_fourier}. Subsequently, we perform a convolution operation with a kernel $\mathcal{C}_k$ on $i_g'(x, y)$ to enhance edge and texture details, as described in Equation~\ref{eq:conv}, where $\ast$ represents the convolution operation.
The kernel $\mathcal{C}_k$ is derived from the Laplacian operator with directional sensitivity.
\begin{equation}\label{eq:inverse_fourier}
i_g'(x, y) = \mathcal{FT}^{-1}\big(f_g'(x_f, y_f)\big)
\end{equation}  

\begin{equation}\label{eq:conv}
\begin{aligned}
C_k=\begin{bmatrix}
-1 & -1 & 1 \\
-1 & 8 & -1 \\
1 & -1 & -1
\end{bmatrix} \\
l_g(x, y) = \mathcal{C}_k \ast i_g'(x, y)
\end{aligned}
\end{equation}

Subsequently, we perform threshold-based binarization operation $\mathcal{T}$ on the image $l_g(x, y)$, converting pixel values within the specified threshold range [$t_l$, $t_h$] to 255:

\begin{equation}
l_g^t(x, y) = \mathcal{T}_{[{t_l},{t_h}]}(l_g(x, y))
\end{equation}  

Finally, we merge the Laplace-convolved output as the fourth channel with the original image, shown in Equation~\ref{eq:merge}.  

\begin{figure}[H]
\centering\includegraphics[width=0.7\textwidth]{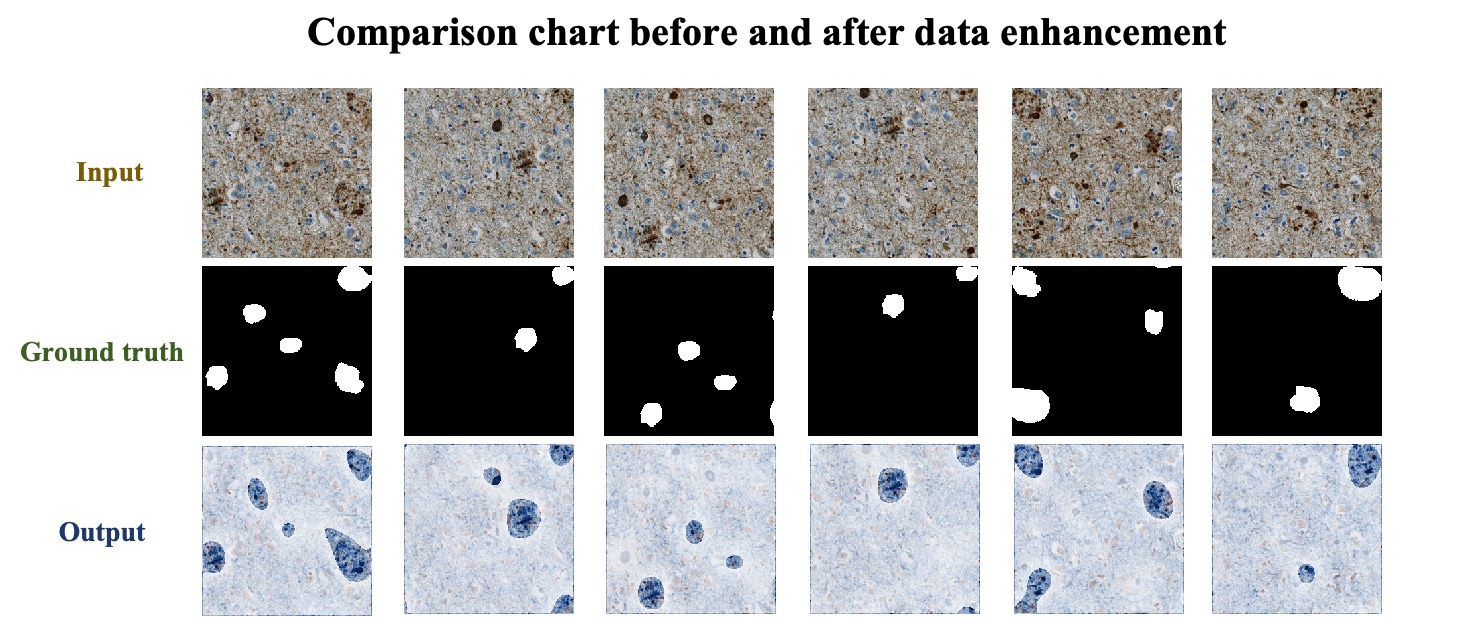}
\caption{The enhanced images after the frequency domain enhancement operation.}
\label{fig:enhance_example}
\end{figure}

The examples of the enhanced images can be seen in Figure~\ref{fig:enhance_example}. As we have observed, the enhanced images effectively highlight the target segment regions.
\begin{equation}\label{eq:merge}
i_f(x, y) = \big(i_r(x, y), i_g(x, y), i_b(x, y), l_g^t(x, y)\big)
\end{equation}

\subsection{Deep learning models for benchmarking}

In our experiments, we evaluated the performance of five deep learning models using four types of color-normalized images, with each type tested on both enhanced and original versions of the images. The experimental methods introduced as follow:
\begin{itemize}
  \item UNet~\cite{ronneberger2015u}: UNet is a classic Convolutional Neural Network (CNN) architecture with a symmetrical encoder-decoder structure. The encoder progressively extracts features, while the decoder progressively restores spatial information. This is complemented by skip connections that incorporate high-resolution features from the encoder, preserving detailed image information. Thanks to its symmetrical encoder-decoder structure and skip connections, UNet effectively retains crucial spatial information and details, making it suitable for processing medical images related to Alzheimer's neuritic plaques.

  \item Attention UNet~\cite{oktay2018attention}: Attention UNet incorporates Attention Gates (AGs) into the classic UNet structure, enabling the model to automatically focus on target structures (such as organs of varying shapes and sizes) while ignoring irrelevant areas. AGs enhance the model's sensitivity and prediction accuracy by suppressing irrelevant areas and highlighting significant features. The introduction of attention mechanisms allows for more precise focus on areas of interest, enhancing the model’s ability to recognize details, which is suitable for processing medical images related to Alzheimer's neuritic plaques.
  
  \item Dual-UNet~\cite{zhou2022generalizable}: Dual-UNet combines dual regularization and style enhancement techniques, simulating the differences between source and target domains to enable the model to handle unseen target domains. The model uses separate batch normalization layers to maintain information from different style images and selects the most suitable path during inference through a style selection module. Its strong generalization capabilities make it particularly applicable to handling images from different patients and different staining methods in this study.

  \item DynUNet~\cite{cardoso2022monai}: DynUNet, a dynamic U-Net implementation within the MONAI framework, supports residual connections, anisotropic convolutional kernels, and deep supervision. This flexibility allows the model to handle medical images of varying scales and shapes more effectively, making it well-suited for applications related to Alzheimer's neuritic plaques.

  \item Swin UNet~\cite{cao2022swinunet}: Swin UNet is a purely encoder-decoder structure based on the Transformer, with skip connections. This model uses hierarchical Swin Transformers as the encoder to learn semantic features from local to global scales, while upsampling through patch expansion layers to restore the spatial resolution of feature maps. The Transformer-based structure provides deep feature learning capabilities from local to global scales, making the model more effective in handling large-scale and high-resolution medical images, and particularly well-suited for processing the extensive medical imagery in this study.

\end{itemize}
\section{Experimental design}

\subsection{Dataset}
The whole-slide image dataset we proposed in this study (ADNP-15) was sourced from the French National Brain Bank (Neuro-CEB), comprising frontal cortex tissue samples from 15 Alzheimer's Disease patients. These samples were collected with written informed consent from the patients or their relatives, and all experimental procedures were approved by the institutional review board, adhering to relevant declarations and regulations concerning human rights. This research is a retrospective analysis aimed at conducting in-depth studies on existing pathological data. The included samples all met the standard conditions for AT8 antibody staining and were scanned using Hamamatsu NanoZoomer 2.0-RS or NanoZoomer S60 scanners. Samples with poor tissue preservation or inadequate staining were excluded. The selected samples showcased varying degrees of tau protein pathology, staining quality, and preservation conditions, ensuring representativeness and diversity of the samples. All scans were performed using Hamamatsu (NanoZoomer series) devices, and stained with AT8 antibodies produced by Thermo Fisher Scientific, USA.

To ensure the accuracy of model training and evaluation, precise manual annotations of the neuritic plaques were performed on each whole-slide image, serving as the ground truth benchmark. Annotations for 4000 plaques in the dataset were completed by experts with many years of experience in neuropathology, ensuring the accuracy and consistency of the annotations. These experts conducted multiple reviews and discussions to meticulously mark the position and boundaries of the plaques, ultimately saving the annotations in XML format for use as a reference in model training. The annotation benchmark was defined based on the morphological characteristics of tau protein plaques observed in the pathological images stained with AT8 antibody under a microscope. These annotations accurately distinguished the subtle differences between neuritic plaques and the surrounding gray matter, offering high reliability and reproducibility. The manual expert annotations were selected as the ground truth benchmark because they accurately capture the complex features and boundary information of the lesions, particularly showing advantages in cases with significant morphological and density changes.

\subsubsection{Data pre-processing}
We extracted image patches of 256$\times$256 pixels from whole-slide images (WSIs) to capture the plaque regions and their surrounding context. A region-guided sampling method was employed to generate these image patches, ensuring that each patch contained the target plaque and its neighborhood background, thereby comprehensively reflecting the relationship between the plaque and adjacent tissues. The contextual information provided by this method is crucial for accurately segmenting neuritic plaques, as these plaques often appear sparse and diffuse within the background gray matter. The design of 256$\times$256 pixel patches provided the model with rich background information, aiding in understanding the spatial distribution and neighborhood characteristics of plaques within the tissue. This context is key to enhancing the model's ability to discern subtle differences between target and non-target structures. Moreover, to reduce the model's dependence on the central position of plaques, a region of interest (ROI)-based translation data enhancement strategy was implemented following our previous study. By shifting the annotated area of each image patch to the four corners, a fourfold data enhancement was achieved, increasing the diversity of plaque positions in different spatial locations and thereby enhancing the model's robustness when dealing with densely distributed plaques.

The ADNP-15 dataset, containing 15 subjects (25,224 image patches, with an average of 1681.1$\pm$1916.05 images per subject), was split into a training set with 12 subjects (18,539 images, averaging 1544.92$\pm$2077.52 images per subject) and a testing set with 3 subjects (6,685 images, averaging 2228.33$\pm$671.05 images per subject), respectively. This split was performed at the patient level to prevent data leakage, as shown in Table~\ref{tab:dataset_distribution}.

\begin{table}[ht]
\caption{Dataset distribution}
\centering
\begin{tabular}{|c|c|c|c|}
\hline
Dataset & Subjects & Patches\\
\hline
Train & 12 & 18539 \\
\hline
Test & 3 & 6685 \\
\hline
Total & 15 & 25224\\
\hline
\end{tabular}
\label{tab:dataset_distribution}
\end{table}

\subsubsection{Stain normalization}

\begin{figure}[H]
\centering
\includegraphics[width=0.6\textwidth]{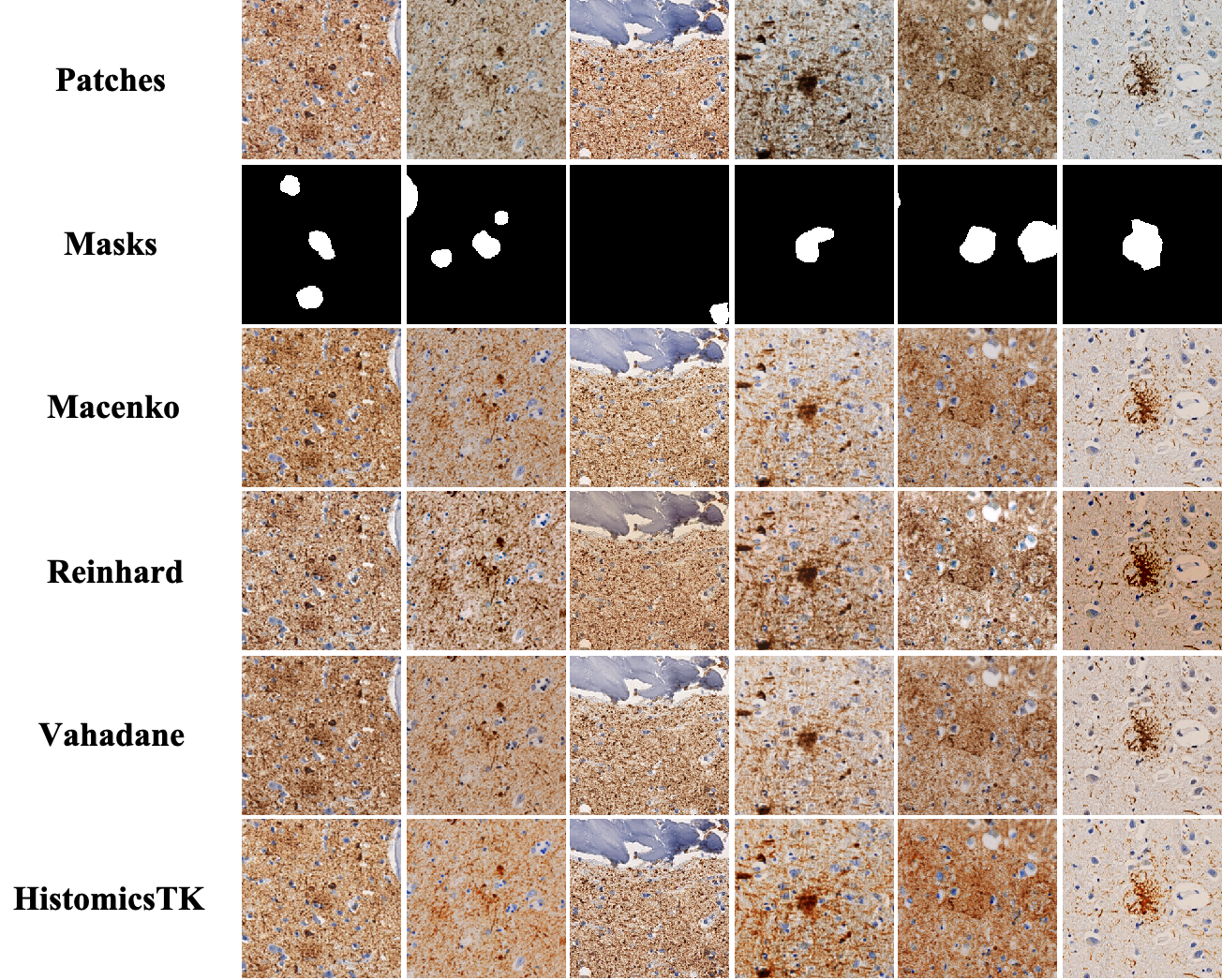}
\caption{Examples of different stain normalization techniques applied to pathological patch images and their corresponding annotated masks.}
\label{fig:stains}
\end{figure}

In medical image analysis, stain normalization is a crucial pre-processing step designed to reduce inconsistencies in image color distributions caused by differences in staining agents, variability in staining processes, and characteristics of the scanning equipment. These color inconsistencies can significantly affect the performance of deep learning-based models, leading to unstable results across different datasets or across multi-center data, thereby limiting the generalization ability of the models. By mapping image color distributions to a unified style, stain normalization not only enhances model robustness but also reduces biases between datasets, enabling models to learn more stable and universal features to maintain excellent performance across different application scenarios. Additionally, stain normalization plays a key role in multi-center studies, facilitating data sharing and model reuse, and promoting collaborative research across institutions. In pathological diagnosis, stain normalization can effectively reduce the impact of staining quality on diagnostic outcomes, enhancing the standardization and consistency of pathological diagnoses, which is crucial for improving the accuracy and reliability of diagnoses~\cite{hoque2024stain}. This study summarizes four stain normalization methods, including Macenko\cite{macenko2009normalizing}, Reinhard\cite{reinhard2001color}, Vahadane\cite{vahadane2016structure}, and HistomicsTK, each with its own characteristics in addressing staining differences. Figure~\ref{fig:stains} displays a series of pathology images processed using different stain normalization methods and their corresponding annotated masks.

\begin{itemize}
  \item Macenko~\cite{macenko2009normalizing}: The Macenko method is based on Optical Density (OD) space for stain normalization. It utilizes Singular Value Decomposition (SVD) to extract staining vectors, thereby deconvolving the image color components to obtain independent concentrations of each stain. This method is highly automated and suitable for the normalization of large-scale tissue slides, especially excelling in quantitative analysis.
  
  \item Reinhard~\cite{reinhard2001color}: The Reinhard method achieves color normalization by transforming the source image into a decorrelated color space called \(l\alpha\beta\). In the \(l\alpha\beta\) color space, the correlation between different channels is minimized. This allows for more independent color adjustments, thus reducing the artifacts that may occur during cross - channel adjustments.

  \item Vahadane~\cite{vahadane2016structure}: The Vahadane method employs Sparse Non-negative Matrix Factorization (SNMF) to separate and standardize stains, aiming to preserve the structural information of tissue slides. It first decomposes the image into stain density maps and alters the color of the source image while maintaining its structure, using the staining basis of the target image. The Vahadane method is particularly suitable for samples with complex tissue structures, as it preserves the details of biological structures and significantly improves the accuracy of normalization.
  
  \item HistomicsTK: This Python-based library does not act as a standalone color normalization tool. Instead, it provides tools for stain normalization via color deconvolution. In this process, the staining matrix of a source image is estimated and then adjusted to match the staining style of a target image. By separating an image into its individual stain components (for example, hematoxylin and eosin), each component can be normalized more precisely, resulting in more consistent and accurate outcomes.
  In this article, we use the term HistomicsTK dataset to refer to the processed dataset obtained through the library’s color deconvolution method.
  
\end{itemize}

\subsection{Evaluation metrics}
To comprehensively evaluate the performance of different models in the task of segmenting neuritic plaques, this study employs a variety of evaluation metrics, specifically including the Dice score, Mean Absolute Surface Distance (MASD), F1-score, and the 95\% bootstrap confidence intervals for these metrics.
The MASD measures the boundary distance between the predicted results and true annotations, assessing the geometric precision of the segmentation. Lower MASD values indicate that the predicted boundaries are closer to the true boundaries, thus higher geometric accuracy.
The F1-score is calculated at the instance level for 2D images, treating each connected component in the segmentation as an individual entity. Segmentation results are first decomposed into separate components, and the Intersection over Union (IoU) is computed between predicted and ground truth components. A component is classified as a true positive (TP) if its IoU with any ground truth component exceeds a user-defined threshold. False positives (FP) occur when predicted components lack corresponding ground truth matches, while false negatives (FN) arise when ground truth components lack predicted matches. The F1-score is then derived from the TP, FP, and FN values, offering a robust measure of instance-level performance. This method evaluates discrete instances rather than pixel-level accuracy, making it well-suited for tasks like object detection and instance segmentation.


\subsection{Implementation details}

The experiments were conducted on an NVIDIA GeForce RTX 4080 GPU with 12 GB of memory, utilizing model architectures from the MONAI framework~\cite{cardoso2022monai}. A fixed learning rate of 0.001 was applied across all experiments, with the Adam optimizer initialized using $\beta_1=0.9$ and $\beta_2=0.999$. The batch size was set to 16.
To address class imbalance and improve segmentation accuracy, the loss function combined Binary Cross Entropy (BCE) and Dice loss. Each deep learning model was trained independently on datasets normalized using four different stain normalization methods. The experimental performance was then evaluated using images generated through the proposed enhancement method.
For the calculation of the instance-level F1-score, the Intersection over Union (IoU) threshold was set to 0.1.
We made the repository\footnote{\url{https://github.com/chenxizhao666/Chenxi_Zhao_AD_pathology_slides_segmentation-3/}} and research data\footnote{\url{https://zenodo.org/records/15039035/}} publicly available.

\section{Results}
\begin{figure}[H]
\centering
\includegraphics[width=0.8\textwidth]{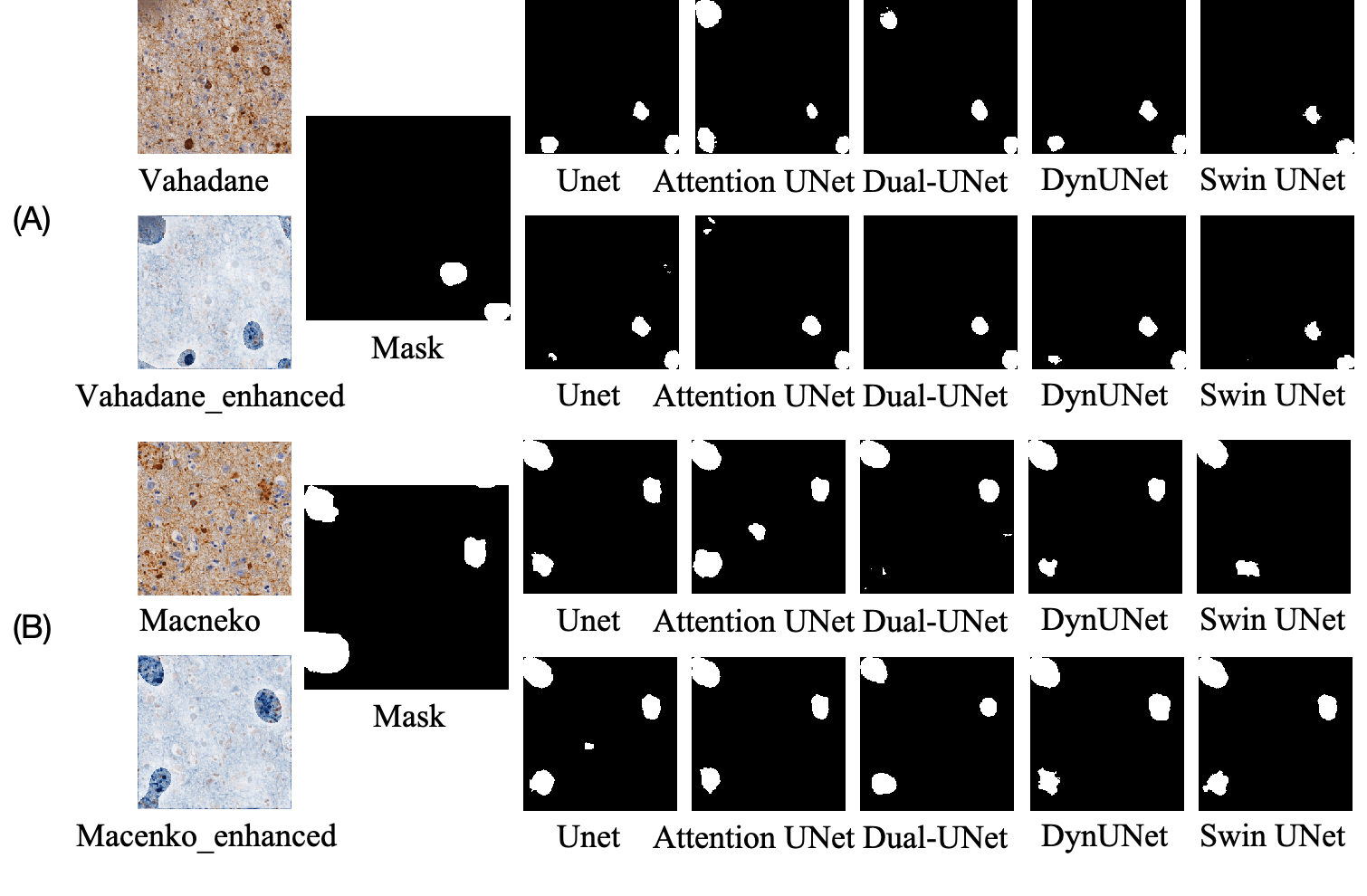}
\caption{Comparison of annotated masks (neuritic plaque) and segmentation results for images processed using four different normalization methods (including images after enhancement), obtained from five deep learning models. }
\label{fig:results}
\end{figure}

\begin{figure}[H]
\centering
\includegraphics[width=0.8\textwidth]{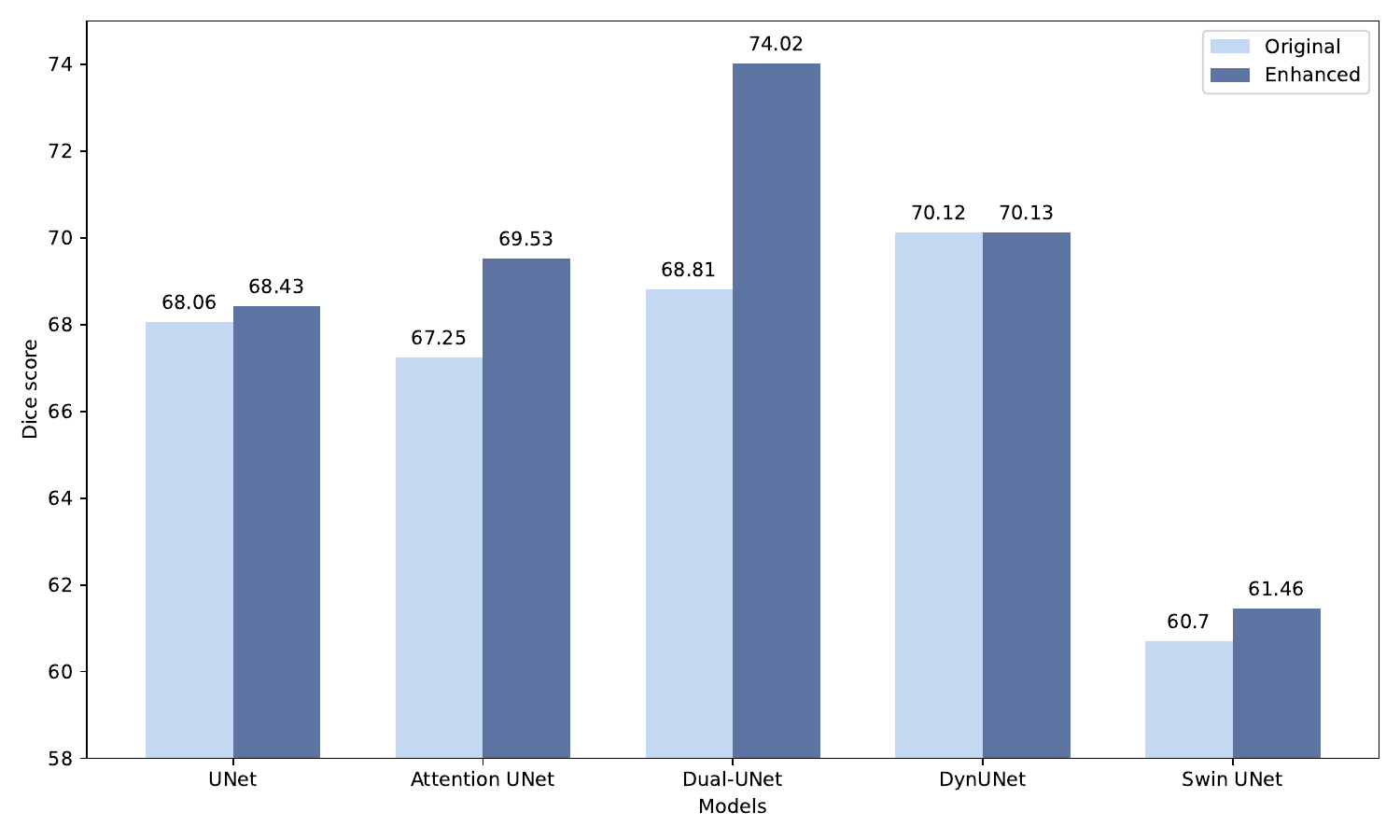}
\caption{Comparison of Dice score for Best Color Normalization Methods Before and After Data Enhancement.}
\label{fig:dice_bar_chart}
\end{figure}

\begin{figure}[H]
\centering
\includegraphics[width=0.9\textwidth]{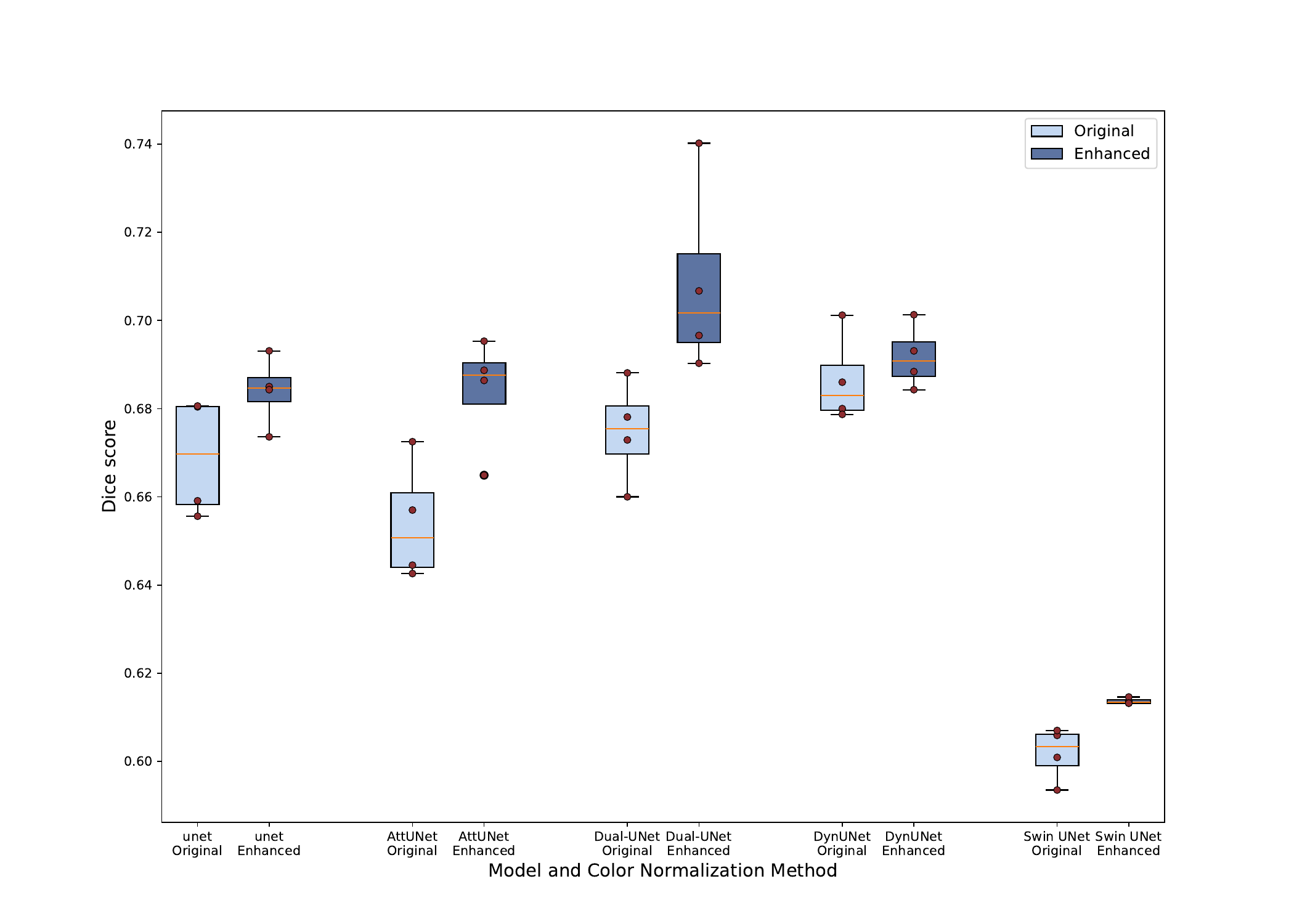}
\caption{Distribution of mean Dice scores for the test set, showing the performance of five deep learning methods on four datasets processed with different normalization techniques and their enhanced versions. Every box is bounded by the lower quartile (Q1) and the upper quartile (Q3), with the centre line representing the median. Whiskers extend from the box to the furthest data point within 1.5x the interquartile range of the box.}
\label{fig:dice_box_chart}
\end{figure}

The experimental results for the neuritic plaque segmentation task, evaluated using five different deep learning models across four stain normalization methods (including the original images and their enhanced versions), are presented in Table~\ref{tab:result:ori} and Table~\ref{tab:result:enhance} separately.
The bar chart of Dice scores is presented in Figure~\ref{fig:dice_bar_chart}, while Figure~\ref{fig:dice_box_chart} illustrates the Dice score distribution of our model across the original images and their enhanced versions. The distribution of MASD and F1-score can be seen in Figure~\ref{fig:masd_box_chart} and Figure~\ref{fig:f1_box_chart} separately. Example segmentation results from different models are shown in Figure~\ref{fig:results}.

\begin{table}[H]
\centering
\caption{Performance of five deep learning methods on four stain-normalized datasets. Results are reported as mean values with their 95\% bootstrap confidence intervals. The F1-score is in instance level (IoU threshold = 0.1).}
\label{tab:result:ori}
\resizebox{1\linewidth}{!}{
\begin{tabular}{|l|l|l|l|l|l|l|} 
\hline
Normalization          & Evaluation metrics & \multicolumn{1}{c|}{Unet} & \multicolumn{1}{c|}{Attention UNet} & \multicolumn{1}{c|}{Dual-UNet} & \multicolumn{1}{c|}{DynUNet} & \multicolumn{1}{c|}{Swin UNet}  \\ 
\hline
\multirow{3}{*}{Macenko}     & Dice score (\%)          & \textbf{68.04[68.03,68.06]}        & 65.70[65.69,65.71]                  & 67.29[66.83,67.74]             & 68.00[67.98,68.01]           & 59.35[58.86,59.86]              \\ 
\cline{2-7}
                             & MASD               & 29.17[28.46,29.89]        & 27.60[26.90,28.32]                  & 30.11[29.24,31.01]             & \textbf{27.77[27.04,28.46]}           & 30.23[29.50,30.98]              \\ 
\cline{2-7}
                             & F1-score (\%)      & 65.29[64.67，65.91]        & \textbf{72.84[72.27,73.42]}                  & 63.56[63.06,64.06]             & 72.62[72.07,73.16]           & 46.41[45.82,47.00]              \\ 
\hline
\multirow{3}{*}{Reinhard}    & Dice score (\%)          & 65.91[65.87,65.94]        & 67.25[67.24,67.26]                  & 67.81[67.34,68.27]             & \textbf{67.87[67.85,67.89]}            & 60.59[60.10,61.09]              \\ 
\cline{2-7}
                             & MASD               & 24.77[24.04,25.58]        & 26.10[25.43,26.85]                  & 26.54[25.72,27.40]             & \textbf{24.66[23.90,25.38]}           & 29.80[29.10.30.51]              \\ 
\cline{2-7}
                             & F1-score (\%)      & 68.12[67.49,68.74]        & 70.58[69.98,71.18]                  & 68.73[68.24,69.22]             & \textbf{75.42[74.87,75.97]}           & 46.70[46.09,47.31]              \\ 
\hline
\multirow{3}{*}{Vahadane}    & Dice score (\%)          & 68.06[68.04,68.08]        & 64.26[64.25,64.27]                  & 66.00[65.53,66.45]             & \textbf{70.12[70.10,70.14]}           & 60.09[59.60,60.59]              \\ 
\cline{2-7}
                             & MASD               & 29.38[28.71,30.04]        & \textbf{26.62[25.95,27.31]}                  & 32.76[31.83,33.71]             & 27.01[26.27,27.70]           & 30.15[29.43,30.86]              \\ 
\cline{2-7}
                             & F1-score (\%)      & 68.34[67.77,68.92]        & 73.10[72.54,73.66]                  & 62.08[61.56,62.61]             & \textbf{75.23[74.68,75.78]}           & 46.44[45.87,47.02]              \\ 
\hline
\multirow{3}{*}{HistonicsTK} & Dice score (\%)          & 65.56[65.52,65.59]        & 64.45[64.44,64.46]                  & \textbf{68.81[68.37,69.23]}             & 68.60[68.58,68.62]           & 60.70[60.24,61.18]              \\ 
\cline{2-7}
                             & MASD               & 26.78[26.08,27.50]        & \textbf{26.02[25.33,26.72]}                  & 29.22[28.39,30.07]             & 26.86[26.12,27.55]           & 31.06[30.32,31.80]              \\ 
\cline{2-7}
                             & F1-score (\%)      & 65.24[64.61,65.87]        & 63.74[63.09,64.39]                  & 68.34[67.59,69.07]             & \textbf{72.34[71.76,72.92]}           & 46.24[45.63,46.85]              \\ 
\hline
\end{tabular}
}
\end{table}

\begin{table}[H]
\centering
\caption{Performance of five deep learning methods on four enhanced stain-normalized datasets generated using the proposed method. Results are reported as mean values with their 95\% bootstrap confidence intervals. The F1-score is in instance level (IoU threshold = 0.1).}
\resizebox{1\linewidth}{!}{
\begin{tabular}{|l|l|l|l|l|l|l|} 
\hline
Normalization          & Evaluation metrics & \multicolumn{1}{c|}{Unet} & \multicolumn{1}{c|}{Attention UNet} & \multicolumn{1}{c|}{Dual-UNet} & \multicolumn{1}{c|}{DynUNet} & \multicolumn{1}{c|}{Swin UNet}  \\ 
\hline
\multirow{3}{*}{Macenko-en}     & Dice score(\%)           & 69.31[69.29,69.34]        & 66.49[66.48,66.50]                  & \textbf{69.66[69.29,70.04]}             & 68.84[68.82,68.86]           & 61.37[60.86,61.85]              \\ 
\cline{2-7}
                             & MASD               & 29.51[28.87,30.22]        & 25.72[25.01,26.46]                  & \textbf{24.44[23.72,25.17]}             & 26.07[25.32,26.76]           & 27.45[26.76,28.14]              \\ 
\cline{2-7}
                             & F1-score(\%)       & 65.88[65.29,66.48]        & 72.89[72.33,73.45]                  & 64.07[63.41,64.73]              & \textbf{76.23[75.71,76.76]}           & 49.44[48.85,50.05]              \\ 
\hline
\multirow{3}{*}{Reinhard-en}    & Dice score(\%)           & 67.36[67.34,67.39]        & 69.53[69.51,69.56]                  & \textbf{70.67[70.25,71.08]}             & 68.43[68.41,68.46]           & 61.46[60.96,61.95]              \\ 
\cline{2-7}
                             & MASD               & 26.29[25.61,26.99]        & 27.61[26.88,28.29]                  & \textbf{18.41[17.73,19.13]}             & 25.58[24.83,26.29]           & 28.89[28.17,29.65]              \\ 
\cline{2-7}
                             & F1-score(\%)       & 68.47[67.88,69.07]        & 71.09[70.51,71.67]                  & 67.95[67.31,68.58]             & \textbf{76.74[76.20,77.29]}           & 47.58[46.95,48.21]              \\ 
\hline
\multirow{3}{*}{Vahadane-en}    & Dice score(\%)           & 68.50[68.49,68.52]        & 68.87[68.86,68.87]                  & 69.03[68.58,69.46]             & \textbf{69.31[69.29,69.33]}           & 61.32[60.86,61.80]              \\ 
\cline{2-7}
                             & MASD               & 29.11[28.42,29.79]        & 27.80[27.10,28.54]                  & \textbf{21.75[21.03,22.48]}             & 25.56[24.81,26.26]           & 29.18[28.45,29.86]              \\ 
\cline{2-7}
                             & F1-score(\%)       &69.89[69.31,70.48]        & 74.34[73.77,74.91]                  & 62.93[61.98,63.87]             & \textbf{74.81[74.23,85.38]}           & 48.30[47.71,48.88]              \\ 
\hline
\multirow{3}{*}{HistonicsTK-en} & Dice score(\%)           & 68.43[68.41,68.45]        & 68.64[68.61,68.67]                  & \textbf{74.02[73.66,74.38]}             & 70.13[70.11,70.15]           & 61.32[60.84,61.60]              \\ 
\cline{2-7}
                             & MASD               & 26.38[25.69,27.03]        & 27.33[26.60,28.03]                  & \textbf{20.08[19.41,20.76]}             & 27.39[26.66,28.08]           & 30.36[29.63,31.09]              \\ 
\cline{2-7}
                             & F1-score(\%)       & 70.61[70.03,71.21]        & 65.36[64.71,66.02]                  & 69.73[69.24,70.42]             & \textbf{75.14[74.60,75.70]}           & 45.35[44.75,45.94]               \\ 
\hline
\end{tabular}
}
\label{tab:result:enhance}
\end{table}
Among the different stain normalization methods, DynUNet consistently demonstrated strong performance. For the Macenko-normalized dataset, DynUNet achieved a high Dice score of 68.00\%, while Attention U-Net and DynUNet obtained the highest F1-scores at 72.84\% and 72.62\%, respectively. However, Swin UNet showed the lowest performance across all metrics. In the Reinhard-normalized dataset, DynUNet achieved the best results, with an F1-score of 75.42\% and a Dice score of 67.87\%. Swin UNet again had the lowest Dice score(60.59\%) and F1-score (46.70\%).  
For the Vahadane-normalized dataset, DynUNet continued to perform best, reaching a Dice score of 70.12\% and an F1-score of 75.23\%. In contrast, Dual-UNet had the lowest F1-score at 62.08\%. In the HistomicsTK-normalized dataset, Dual-UNet achieved the highest Dice score (68.81\%) and F1-score (68.34\%), with DynUNet following closely with an F1-score of 72.34\%. Swin UNet consistently exhibited the lowest performance, with an F1-score of 46.24\%.  
Overall, DynUNet consistently achieved high F1-scores and Dice score across most stain-normalized datasets, indicating its robustness for this task. In contrast, Swin UNet underperformed in all cases, highlighting its limited effectiveness. While Attention U-Net showed strong performance on some datasets, it did not consistently surpass DynUNet. These findings suggest that DynUNet is a strong candidate for stain-normalized medical image segmentation, with the choice of normalization method significantly impacting segmentation performance.  
Among the four stain normalization methods, Vahadane and Reinhard normalization produced the highest segmentation performance overall, as indicated by the higher Dice score and F1-scores across multiple deep learning models. Specifically, DynUNet achieved its best results with Vahadane normalization (Dice score: 70.12\%, F1-score: 75.23\%), while the highest F1-score across all methods was observed with Reinhard normalization (DynUNet, 75.42\%). In contrast, Macenko and HistomicsTK normalization resulted in more variable performance, with larger discrepancies between different models. Notably, Swin UNet exhibited consistently poor performance across all normalization techniques, suggesting that it is less robust to stain variation. These results highlight the importance of selecting an appropriate stain normalization method, with Vahadane and Reinhard being the most effective for improving segmentation accuracy.

After applying image enhancement, the segmentation performance improved across most deep learning models and stain normalization methods. For the Macenko-enhanced dataset, DynUNet achieved the highest F1-score (76.23\%), showing a notable improvement compared to the original dataset (72.62\%). Similarly, the highest Dice score was observed for Dual-UNet (69.66\%), an increase from 67.29\% before enhancement. Swin UNet continued to show the lowest performance, but its F1-score improved from 46.41\% to 49.44\%.
In the Reinhard-enhanced dataset, DynUNet again demonstrated the highest F1-score (76.74\%), surpassing its previous performance (75.42\%). Dual-UNet achieved the highest Dice score (70.67\%), an increase from 67.81\%. The MASD values were also reduced for several models, suggesting an improvement in segmentation accuracy. Swin UNet, however, continued to lag behind, with its F1-score only reaching 47.58\%.
For the Vahadane-enhanced dataset, DynUNet achieved an F1-score of 74.81\%, which remained slightly lower than its performance on the Reinhard-enhanced dataset. Dual-UNet and U-Net exhibited improvements in Dice scores, reaching 69.03\% and 68.50\%, respectively. Swin UNet showed a small improvement in F1-score but still underperformed relative to the other models.
In the HistomicsTK-enhanced dataset, Dual-UNet outperformed other models in Dice score (74.02\%) and maintained an F1-score of 69.73\%, an increase from 68.34\%. DynUNet continued to perform well, reaching an F1-score of 75.14\%, compared to 72.34\% in the original dataset. Swin UNet again exhibited the weakest performance, with its F1-score decreasing slightly to 45.35\%.
Overall, image enhancement led to consistent improvements in Dice score and F1-scores across all deep learning models and normalization methods, particularly benefiting DynUNet and Dual-UNet. Among the stain normalization methods, Reinhard-enhanced and Vahadane-enhanced datasets yielded the most significant performance gains, with Reinhard-enhanced data producing the highest F1-score (76.74\%, DynUNet). Additionally, MASD values were generally reduced, indicating more precise segmentation boundaries.
The enhancement had a relatively smaller impact on Swin UNet, which still exhibited the lowest performance across all datasets. However, slight improvements in Dice score and F1-score suggest that the proposed enhancement method helped mitigate some of its weaknesses. Notably, the performance gap between DynUNet and other models widened after enhancement, reinforcing its robustness in stain-normalized image segmentation.

In summary, the proposed image enhancement strategy effectively enhances segmentation performance, particularly benefiting DynUNet and Dual-UNet, with the best results observed on Reinhard-enhanced and Vahadane-enhanced datasets. This highlights the importance of both stain normalization and data enhancement in improving whole slice image segmentation accuracy.

\section{Conclusion}

In this study, we open-sourced a post mortem human brain WSI dataset from patients with Alzheimer’s disease for neuritic plaque segmentation. Additionally, we systematically evaluated the performance of four stain-normalized datasets across five deep learning segmentation models. To further improve segmentation accuracy, we proposed a novel image enhancement method. Experimental results demonstrated that the proposed enhancement method significantly improved segmentation accuracy, particularly in handling complex tissue structures. Among the evaluated deep learning models, Dual-UNet with enhanced images achieved the highest performance under the color deconvolution normalization.  
Future research could focus on developing normalization-free methods to enhance generalizability and efficiency. Our study establishes a new benchmark in this field, and the open-source dataset can serve as a valuable resource for advancing research in Alzheimer’s disease analysis.

\section{Acknowledgement}
This research was supported by Mr Jean-Paul Baudecroux and The Big Brain Theory Program - Paris Brain Institute (ICM). The human samples were obtained from the Neuro-CEB brain bank (\url{https://www.neuroceb.org/en/}) (BRIF Number 0033-00011),  partly funded by the patients’ associations  ARSEP, ARSLA, “Connaître les Syndromes Cérébelleux,” France-DFT, France Parkinson and by Vaincre Alzheimer Foundation, to which we express our gratitude. We are also grateful to the patients and their families.

\noindent This work was granted access to the HPC resources of IDRIS under the allocation 2023-AD011013250 made by GENCI.

\noindent Gabriel Jimenez is associated with the Paris Brain Institute Quantitative Cellular and Molecular Imaging Core Facility (RRID: SCR\_026393).

\bibliography{refs} 
\bibliographystyle{spiebib} 
\newpage
\appendix
\renewcommand\thefigure{S\arabic{figure}} 
\setcounter{figure}{0} \renewcommand\thetable{S\arabic{table}} 
\setcounter{table}{0}   
\section{Performance distribution on other evaluation metrics}

\begin{figure}[!ht]
\centering
\includegraphics[width=0.7\textwidth]{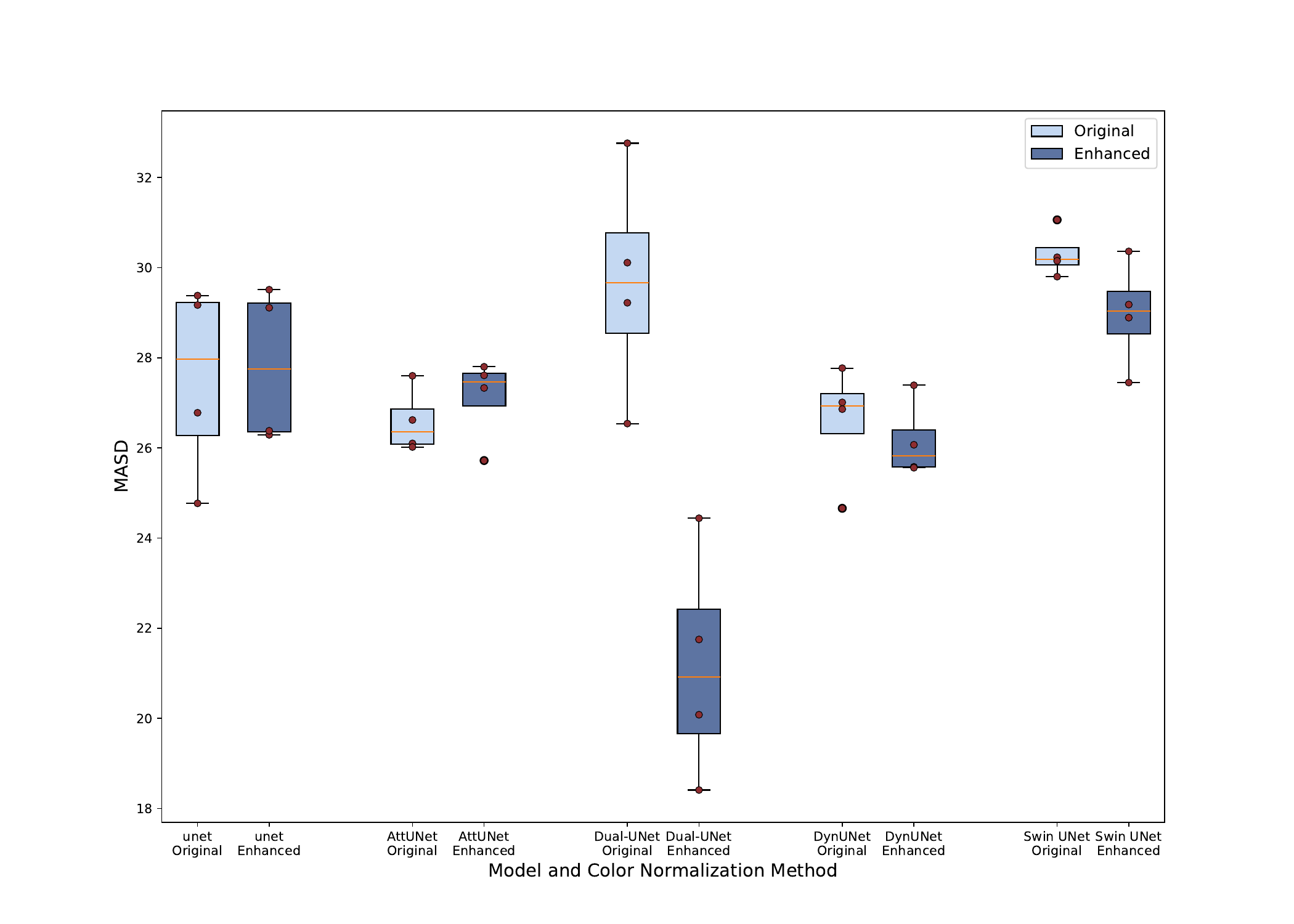}
\caption{MASD comparison of different models across various normalized dataset, including the original images and their enhanced versions.}
\label{fig:masd_box_chart}
\end{figure}

\begin{figure}[!h]
\centering
\includegraphics[width=0.7\textwidth]{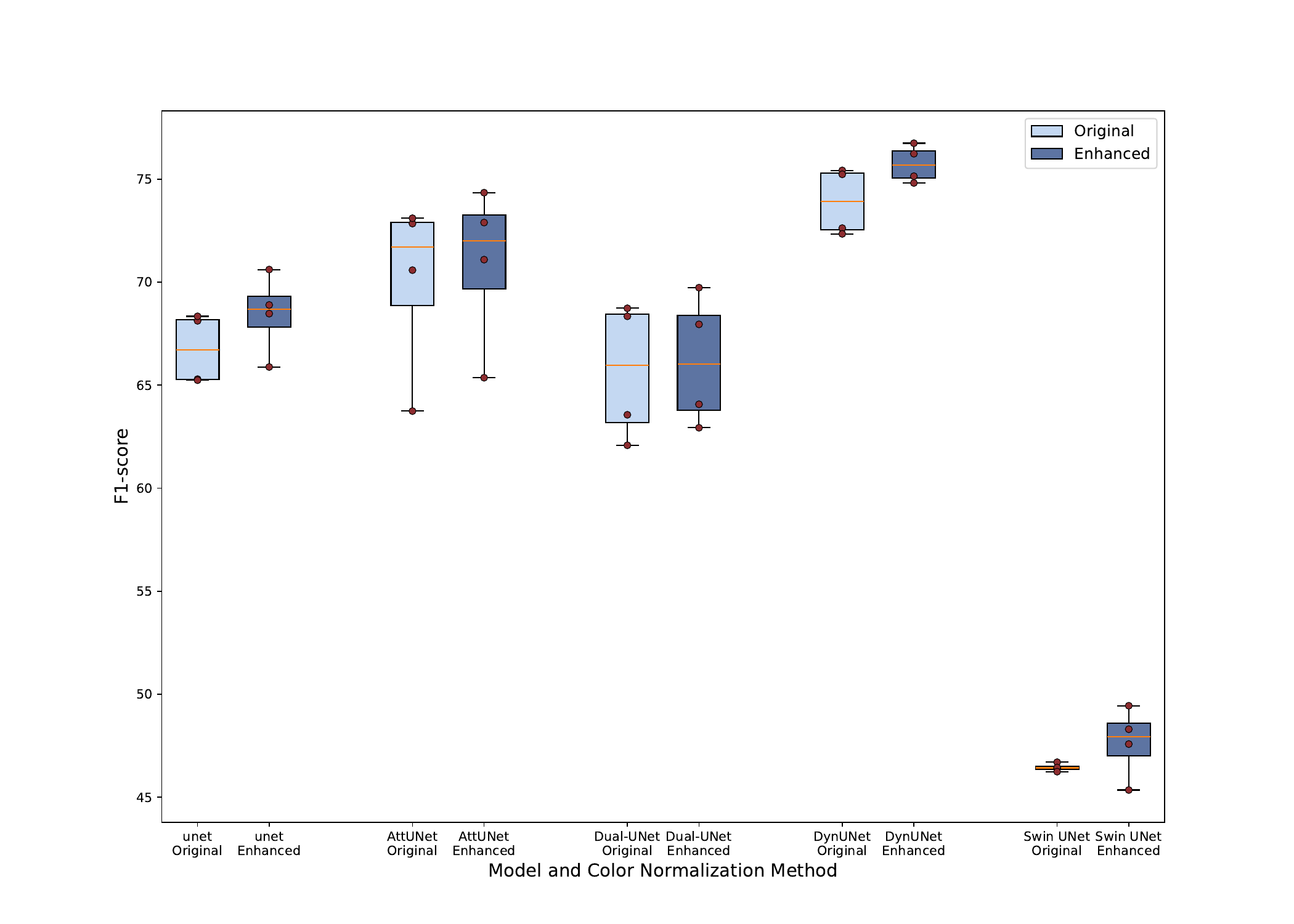}
\caption{F1-score comparison of different models across various normalized dataset, including the original images and their enhanced versions.}
\label{fig:f1_box_chart}
\end{figure}

Figure~\ref{fig:masd_box_chart} and~\ref{fig:f1_box_chart} display the MASD and F1-score distribution for the test set with different normalization. 

\begin{table}[H]
\centering
\caption{The F1-score performance of five deep learning methods on four enhanced stain-normalized datasets generated using the proposed data augmentation method and on datasets without data augmentation. The results are presented as mean values along with their 95\% bootstrap confidence intervals. The F1-score is at the instance level (IoU threshold = 0.5). }
\label{tab:result:comparison}
\resizebox{1\linewidth}{!}{
\begin{tabular}{|l|l|l|l|l|l|} 
\hline
Models                      & Macenko                  & Reinhard                 & Vahadane                 & HistonicsTK              \\ 
\hline
Unet                        & 52.84 [52.15, 53.53]     & 56.47 [55.78, 57.17]     & 56.01 [55.34, 56.69]     & 53.52 [52.80, 54.24]     \\ 
\hline
Unet\_enhanced              & \textbf{53.75 [53.03, 54.47]}     & \textbf{57.89 [57.16, 58.62]}     & \textbf{56.63 [55.94, 57.33] }    & \textbf{57.98 [57.28, 58.68]}     \\ 
\hline
Attention UNet              & 60.27 [59.56, 60.97]     & 58.48 [57.78, 59.18]     & 59.73 [59.03, 60.43]     & 52.05 [51.34, 52.75]     \\ 
\hline
Attention UNet\_enhanced    & \textbf{60.32 [59.61, 61.02]}     & \textbf{59.56 [58.86, 60.26]}     & \textbf{62.03 [61.32, 62.74]}     & \textbf{54.26 [53.53, 54.98]}     \\ 
\hline
Dual-UNet                   & 45.97 [45.24, 46.71]     & 48.00 [47.24, 48.76]     & 45.94 [45.21, 46.67]     & 47.81 [47.11, 48.51]     \\ 
\hline
Dual-UNet\_enhanced         & \textbf{53.20 [52.47, 53.92]}     & \textbf{56.98 [56.24, 57.71]}     & \textbf{50.52 [49.44, 51.59]}     & \textbf{57.03 [56.58, 57.93]}     \\ 
\hline
DynUNet                     & 56.17 [55.46, 56.88]     & 63.28 [62.59, 63.97]     & 62.75 [62.04, 63.47]     & 60.81 [60.11, 61.51]     \\ 
\hline
DynUNet\_enhanced           & \textbf{62.98 [62.29, 63.67]}     & \textbf{63.77 [63.07, 64.47]}     & \textbf{63.01 [62.32, 63.70]}     & \textbf{62.74 [62.05, 63.44]}     \\ 
\hline
Swin UNet                   & 34.67 [34.04, 35.29]     & 35.38 [34.74, 36.02]     & 34.9 [34.29, 35.51]      & \textbf{35.25 [34.62, 35.88]}     \\ 
\hline
Swin UNet\_enhanced         & \textbf{36.87 [36.23, 37.51]}     & \textbf{35.58 [34.94, 36.24]}     & \textbf{36.03 [35.41, 36.65]}     & 34.35 [33.74, 34.95]    \\ 
\hline
\end{tabular}
}
\end{table}

\end{document}